\documentclass[journal]{IEEEtran}

\usepackage{amsmath,amssymb}
\usepackage{tikz}
\usetikzlibrary{positioning,arrows.meta,calc}
\usepackage{booktabs}
\usepackage{array}
\usepackage{url}
\usepackage{balance}

\newcommand{\eg}{e.g.,}
\newcommand{\ie}{i.e.,}
\newcommand{\stepitem}{\textbullet~}

\begin{document}

\title{Context-Aware Trust Management for Autonomous Networks: A Modular Framework toward 6G}

\author{Daniel Cho,
\and Loay Abdelrazek,
\and Rim~El~Malki}


\maketitle

\begin{abstract}
Networks are evolving toward autonomous operation: closed automation loops, increasing levels of autonomy as defined by industry initiatives, and, in the 6G era, AI-native network functions (NFs) that configure, heal, and optimize themselves with minimal human involvement. In such an environment, static, identity-based security decisions are no longer sufficient: an autonomous network can only act as fast and as safely as the trust it places in its own functions, services, and data, so trust must be evaluated continuously, per context, and adapted over time, in line with the continuous-evaluation tenet of the zero-trust architecture. This paper presents a modular framework for context-aware trust management (TM) in autonomous networks. We define the notion of context for the radio access network (RAN) domain of an autonomous network, instantiated on cloud-native distributed units (DUs) and central units (CUs), identifying two main context categories, asset context and services context, the latter subdivided into 3GPP services (the F1, E1, and Xn interfaces and their procedures) and non-3GPP application support services. We identify the conceptual components required to realize context-aware TM, namely context metrics aggregation, trust calculation, trust level thresholding, response strategies, and a knowledge base, and formalize them as an ontology in the form of an explicit graph that captures the dependencies and data flows between components and their inputs. To make the framework actionable, we provide a simple trust evaluation model: per-context trust scores are computed as weighted aggregations of normalized security metrics, combined into an overall NF score, and compared against a dynamic trust level threshold that adapts to performance indicators, impact and risk assessments, and historical scores. We describe the end-to-end sequence flow and three deployment options: centralized in a security orchestrator, distributed across NFs, and realized as an O-RAN rApp/xApp suite feeding the closed automation loops of the autonomous network. The framework complements, rather than replaces, existing security controls and enables granular, non-intrusive, context-dependent decisions in telecom networks, from 5G toward 6G.
\end{abstract}


\section{Introduction}
\IEEEPARstart{M}{obile} networks are on a path toward autonomous operation. Industry initiatives such as the TM Forum's autonomous networks program and ETSI's work on autonomous networks define a maturity ladder from assisted operations to full autonomy, in which the network carries out its own intent-translation, configuration, healing, and optimization in closed automation loops with minimal human involvement~\cite{tmforum,etsian}. The radio access network (RAN) is a central part of this evolution: RAN functions such as the distributed unit (DU) and central unit (CU) are deployed as cloud-native workloads on shared infrastructure, and their management is increasingly delegated to AI- and ML-driven automation, a trend that the 6G generation is expected to take further with AI-native network design. At the same time, advanced persistent threats and supply-chain risks persist, and the attack surface grows with every automated, dynamic, and distributed component.

Autonomy changes the security question. A human operator can weigh whether a function looks healthy before acting on it; an autonomous network must decide by itself, continuously, which functions, services, and data to rely on. Security decisions based solely on identity, static credentials, or perimeter position are therefore inadequate: the network needs a quantifiable, continuously updated measure of how trustworthy each entity is, per context, to drive its closed loops safely.

Trust management (TM) offers a complementary approach: rather than asking only \emph{who} an entity is, a TM system continuously evaluates \emph{how trustworthy} an entity is, based on observations of its behavior, configuration, and environment, and expresses the result as a quantifiable trust score. Continuous trust evaluation is one of the tenets of the zero-trust architecture (ZTA) defined by NIST~\cite{nist800207}. However, existing TM approaches exhibit several shortcomings when applied to autonomous telecom networks: the notion of \emph{context} in the literature is narrow, typically limited to location, requester identity, and roles, which is insufficient to characterize an asset in a telecommunication environment where configuration changes, deployed security controls, emerging threats, topology, and running services all matter; no context-aware TM system has, to the best of our knowledge, been applied to telecom systems, leaving unclear what context-awareness means for a RAN NF and which contextual information can be collected; existing TM frameworks are monolithic rather than modular; trust thresholds are static; and existing systems lack granular, context-dependent response actions, although an asset may require different decisions in different contexts.
This paper addresses these gaps. Its contributions are:
\begin{enumerate}
\item a definition of context for the RAN domain of an autonomous network, organized into two categories, \emph{asset context} and \emph{services context}, the latter covering both 3GPP-defined services and non-3GPP application support services;
\item an identification of the security metrics and measures that can quantify each context with respect to trust;
\item a modular decomposition of a context-aware TM system into five components, formalized as an ontology in the form of an explicit graph that captures components, their inputs, and their dependencies;
\item a simple, formally stated trust evaluation and threshold-adaptation model that makes the framework actionable and explains how scores, historical trends, and thresholds interact;
\item the end-to-end sequence flow of the framework and three deployment options: centralized in a security orchestrator, distributed across NFs, and realized as an O-RAN rApp/xApp suite feeding the closed automation loops.
\end{enumerate}

The remainder of this paper reviews trust, context, and TM (Section~\ref{sec:background}), defines the contexts of a RAN NF and their candidate metrics (Sections~\ref{sec:contexts} and \ref{sec:metrics}), presents the trust evaluation and threshold-adaptation model (Section~\ref{sec:model}), the ontology and sequence flow (Sections~\ref{sec:ontology} and \ref{sec:flow}), and the deployment options, discussion, and conclusion (Sections~\ref{sec:deployment}--\ref{sec:conclusion}).

\section{Background}\label{sec:background}

\subsection{The Notion of Trust}
We adopt the definition of trust as \emph{a characteristic of an entity that indicates its ability to perform certain functions or services correctly, fairly, and impartially, along with the assurance that the entity and its identifier are genuine}~\cite{nistglossary}. A \emph{trustor} (the subject that trusts) holds a belief and confidence that the \emph{trustee} (the target entity) is honest, available, dependable, and secure; trust should therefore not be established on identity alone.

Three properties of trust motivate the design in this paper. First, trust is \emph{dynamic}: in a complex, distributed network such as the RAN, several factors can change the trust of an asset over time, so trust must be continuously monitored, evaluated, and adjusted. Second, trust is \emph{context dependent}: an NF may be trusted in one context (a given interface or service) and not in another. Third, trust can be \emph{direct}, formed from direct observations and interactions between trustor and trustee, or \emph{indirect}, bootstrapped by recommendations from third parties when no such history exists.\subsection{The Notion of Context}
Following NIST's definition of system context as \emph{the specific system elements, boundaries, interconnections, interactions, and operational environment that define a system}~\cite{nistglossaryctx}, we treat context as any information that characterizes the state (when, where, and how) of an entity or of the environment in which the entity operates~\cite{dey}. Context is thus the basic building block of context-aware TM: without a rich definition of what the contexts of an entity are, a TM function cannot be context-aware.
\subsection{Trust Management}
Trust management refers to managing all trust aspects in an environment: defining what trust is, defining the involved entities, trust calculation, trust propagation and aggregation, and decision-making based on trust scores~\cite{iotstudy,cloudsurvey}. A TM system observes the behavior of assets (\eg{} NFs, services, applications, interfaces, and APIs), their compliance with policies, procedures, and expected behavior, and assigns each asset a trust score that denotes how reliable, secure, and dependable it is, so that trusted relationships can be built between assets, untrusted assets identified, and appropriate actions taken.

TM can be centralized or decentralized. In a decentralized design, the entities themselves collect contextual data and evaluate each other; in a distributed, automated environment such as an autonomous network this is problematic, since a compromised evaluating entity may issue malicious trust evaluations and mislead the closed automation loops that rely on them. A centralized design, in which a single security function (\eg{} a security orchestrator such as Ericsson Security Manager, ESM) collects contextual information from many systems and conducts the evaluations, requires only the central entity to be secured and benefits from broader data collection, which improves evaluation accuracy. We therefore assume a centralized TM function, while noting that the modular design of Section~\ref{sec:ontology} also supports hybrid deployments.

\subsection{Trust Management Techniques}
The literature distinguishes several TM techniques~\cite{iotstudy,cloudsurvey}. In \emph{policy-based} TM, trust is evaluated against pre-defined rules, most commonly a minimum trust threshold that must be fulfilled for access to be granted (\eg{} possession of a certificate from a trusted authority); declarative policies suit access control in open distributed systems because they yield Boolean decisions~\cite{bonatti}. In \emph{reputation-based} TM, trust is evaluated from the aggregation of ratings communicated by other trusted entities. In \emph{recommendation-based} TM, entities with prior experience of a service send recommendations to an inquiring entity, explicitly or transitively; a drawback is that recommendations reflect the recommender's own perspective and may be biased. In \emph{prediction-based} TM, entities evaluate each other based on the similarity of their interests and capabilities, which helps avoid risky interactions with untrusted entities.

\emph{Context-aware} TM can combine the techniques above, but its distinguishing aspect is context-awareness itself: the TM function gathers contextual information about the environment and dynamically adapts its behavior based on the context of the NF, application, user, or service being evaluated~\cite{dey}. Since an autonomous network is complex, dynamic, and distributed, the contexts of an NF change over time, and therefore the trust of the NF changes over time as well. A context-aware TM function monitors the RAN domain, specifically the DU and CU, collects contextual information, and enriches it with data from other domains (\eg{} detection functions, user and entity behavior analytics (UEBA), vulnerability management, and network management), improving the accuracy of the trust scores before responding with an appropriate strategy. Closest to our work, the TM framework for containerized workloads in 5G networks of~\cite{meloudi} defines Kubernetes-specific context parameters, and patents have addressed trust scoring from network performance results under static thresholds~\cite{pat1,pat2}; neither covers the richer context space of telecom NFs nor adaptive thresholds, which this paper addresses.
\section{Contexts in the RAN Domain of an Autonomous Network}\label{sec:contexts}

\begin{figure}[t]
\centering
\begin{tikzpicture}[
  box/.style={draw, fill=white, rounded corners=1.5pt, align=center, font=\scriptsize, inner sep=3pt, minimum height=6mm},
  edge/.style={-{Latex[length=1.6mm]}, font=\scriptsize}
]
\node[box] (nf) at (0,0) {RAN NF\\(CU-UP, CU-CP, DU)};
\node[box] (asset) at (-2.1,-1.5) {Asset context};
\node[box] (svc) at (2.1,-1.5) {Services context};
\node[box] (wn) at (-3.1,-3.0) {Worker node};
\node[box] (pod) at (-1.1,-3.0) {Pod};
\node[box] (gpp) at (0.9,-3.0) {3GPP services\\(F1, E1, Xn)};
\node[box] (appss) at (3.2,-3.0) {Non-3GPP\\(AppSS)};
\draw[edge] (nf) -- (asset);
\draw[edge] (nf) -- (svc);
\draw[edge] (asset) -- (wn);
\draw[edge] (asset) -- (pod);
\draw[edge] (svc) -- (gpp);
\draw[edge] (svc) -- (appss);
\end{tikzpicture}
\caption{Context taxonomy of a RAN network function. Asset context covers the hosting infrastructure (worker nodes and pods); services context covers 3GPP-defined interfaces and their procedures, and non-3GPP application support services (AppSS).}
\label{fig:contexts}
\end{figure}
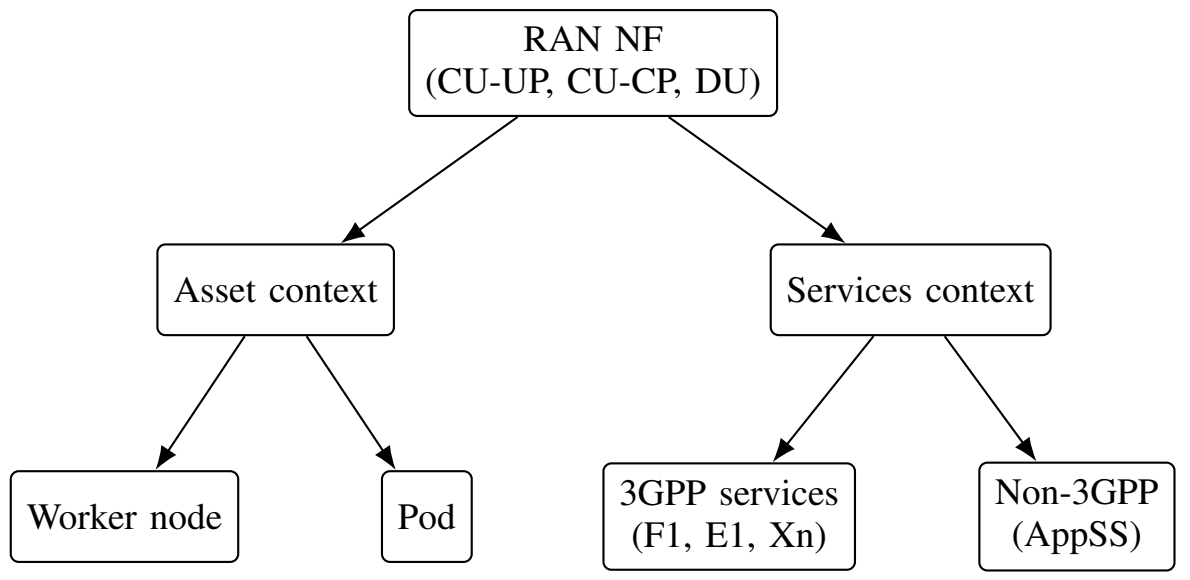

Context is the fundamental building block of context-aware TM. For the RAN domain of an autonomous network, instantiated on cloud-native RAN NFs, we identify two context categories: \emph{asset context} and \emph{services context}, as illustrated in Fig.~\ref{fig:contexts}.

The \emph{asset context} refers to the contexts that exist in the assets themselves: the worker nodes hosting the cluster, the pods hosted on those worker nodes, and the logical NFs CU-UP, CU-CP, and DU. The \emph{services context} refers to the services the NF provides or consumes, in two sub-categories. \emph{3GPP services} are the 3GPP-defined interfaces employed by the RAN NFs, \ie{} F1, E1, and Xn, including the procedures they carry (\eg{} F1 interface management, E1 bearer context management). \emph{Non-3GPP services} are the application support services (AppSS) provided by the application development platform (ADP), such as security services (certificate management, administrator user management, LDAP) and configuration management services.

The motivation for this categorization is that services and functions in the autonomous network are generally implemented as containers, pods, or virtual machines, so a service consumer may trust a service or function but not the underlying platform that runs it, or conversely. Separating asset and services contexts makes this distinction explicit and evaluable: each interface, procedure, and AppSS constitutes an independently evaluable context of the NF, so that at a given point in time one context may be trusted while another is not (\eg{} the F1 interface management procedures may be trusted while the configuration management services are not). This enables granular decision-making and response that affects only a small part of the NF rather than the whole NF, which is particularly valuable for NFs considered critical and for closed automation loops that must act on precise, localized information.

\section{Security Metrics and Measures}\label{sec:metrics}

For a TM function to evaluate the trust of entities through their contexts, each context must itself be evaluated with respect to trust. Quantifiable security metrics are especially suitable here, as they express various security aspects of an entity and its contexts numerically. Table~\ref{tab:metrics} lists four main metrics we identified, with representative measures for each; the measures are examples rather than an exhaustive list, and a detailed investigation of metrics and measurement methods is a separate study.

\begin{table}[t]
\caption{Identified security metrics and representative measures.}
\label{tab:metrics}
\centering
\footnotesize
\begin{tabular}{p{2.45cm}p{5.45cm}}
\toprule
\textbf{Metric} & \textbf{Representative measures} \\
\midrule
Attacks and compromises & total security incident events; ratio of true-positive incidents; ratio of false-positive alerts; suspicious human and non-person-entity behaviors \\
\addlinespace
Hardening and compliance coverage & non-secure or non-compliant configurations w.r.t.\ benchmarks (\eg{} CIS); ratio of patched systems; OS, kernel, container runtime, and Kubernetes versions; number of users with administrator privilege \\
\addlinespace
Vulnerability posture & number of known vulnerabilities; number and ratio of high/critical vulnerabilities; number of exploitable vulnerabilities; patch cadence and time to patch \\
\addlinespace
Attack surface & unidentified devices in the network; number of attack paths w.r.t.\ exploitable vulnerabilities; newly added assets; direct interactions with sensitive resources \\
\bottomrule
\end{tabular}
\end{table}

\section{Trust Evaluation and Threshold Adaptation}\label{sec:model}

This section formalizes the evaluation logic of the framework. The model below is deliberately simple, so that each quantity is interpretable, and any component can later be replaced by a more sophisticated algorithm without changing the framework.

\subsection{Per-Context Trust Scores}Let $n$ denote the network function under evaluation and $\mathcal{C}_n$ its set of contexts (asset and services contexts, Section~\ref{sec:contexts}). For a context $c \in \mathcal{C}_n$, the data correlation component produces a set of normalized metric values
\begin{equation}
\tilde{m}_{c,1}(t), \tilde{m}_{c,2}(t), \dots, \tilde{m}_{c,K_c}(t) \in [0,1],
\label{eq:metrics}
\end{equation}
where $K_c$ is the number of metrics used for context $c$ and $t$ is the evaluation time. Normalization maps each raw measure (\eg{} the ratio of patched systems, the percentage of critical vulnerabilities) onto a common scale in which $1$ is the most favorable value for trust. The trust score of context $c$ is the weighted aggregation
\begin{equation}
T_n(c,t) \;=\; \sum_{k=1}^{K_c} \alpha_{c,k}\, \tilde{m}_{c,k}(t),
\qquad \sum_{k=1}^{K_c} \alpha_{c,k} = 1,
\label{eq:contextscore}
\end{equation}
where $\alpha_{c,k} \geq 0$ is the weight of metric $k$ in context $c$. In words: the trust of a context is a weighted average of its security metrics, so a context is as trustworthy as its weakest relevant aspects, weighted by their importance. The weights can be set by policy or learned; \eg{} exploitable vulnerabilities may be given a higher weight than patch cadence.
\subsection{Overall NF Trust Score}
The overall trust of the NF aggregates its per-context scores:
\begin{equation}
T_n(t) \;=\; \sum_{c \in \mathcal{C}_n} w_c\, T_n(c,t),
\qquad \sum_{c \in \mathcal{C}_n} w_c = 1,
\label{eq:overallscore}
\end{equation}
where $w_c \geq 0$ reflects the criticality of context $c$ to the NF's mission. Keeping the per-context scores $T_n(c,t)$ available, rather than only the aggregate, is what enables granular responses: the decision-making component can act on the specific context whose score is low (Section~\ref{sec:flow}).

\subsection{Historical Scores and Outlier Detection}
Trust evolves over time, and a single score should not be treated in isolation. The knowledge base stores past scores, from which a historical reference is computed over a sliding window $W$:
\begin{equation}
\bar{T}_n(t) \;=\; \frac{1}{|W|} \sum_{s \in W} T_n(s).
\label{eq:history}
\end{equation}
If the current score deviates sharply from this reference, $\lvert T_n(t) - \bar{T}_n(t) \rvert \geq \epsilon$ for a tolerance $\epsilon$, the score is flagged as a probable outlier and re-evaluated before any decision is taken, reducing the risk of both false positives (an abrupt but legitimate change triggering a drastic response) and false negatives (a gradual degradation being accepted as normal).

\subsection{Dynamic Trust Level Threshold}
Each context of each NF has a trust level threshold $\tau_n(c,t)$ against which its score is compared. Unlike static thresholds, $\tau$ is recomputed continuously from requirements stored in the knowledge base: performance indicators and KPIs (\eg{} mean time to detect, mean time to resolve, mean time to acknowledge, mean time between failures), impact and risk assessment inputs (\eg{} human cost, data loss cost, reputation damage, financial damage, system downtime, asset criticality), and the NF's historical scores. A simple adaptation rule, applied after each evaluation cycle, is (writing $T$, $\bar{T}$, and $\tau$ for $T_n(t)$, $\bar{T}_n(t)$, and $\tau_n(t)$):
\begin{equation}
\tau_n(t{+}1) =
\begin{cases}
\min\!\big(\tau_n(t) + \delta,\; 1\big), & T > \tau_n(t) \wedge \bar{T} > \tau_n(t),\\[2pt]
\tau_n(t), & T = \tau_n(t),\\[2pt]
\max\!\big(\tau_n(t) - \delta,\; 0\big), & T < \tau_n(t) \wedge \bar{T} < \tau_n(t),
\end{cases}
\label{eq:threshold}
\end{equation}
where $\delta > 0$ is a small adaptation step and $\bar{T}_n(t)$ is the historical reference of \eqref{eq:history}. The three cases follow the framework's comparison outcomes. If both the current score and the historical reference exceed the threshold, the asset consistently performs better than required, so the threshold can be raised (\eg{} when additional security or network requirement costs are deemed negligible). If the score equals the threshold, it was accurate for the given insights and is left unchanged. If both are below the threshold, the threshold is lowered, conditionally: for a critical NF, an operator may prioritize service continuity over a strict security level, and lowering the threshold lets the NF keep running while its contexts are remediated.
\subsection{Decision Rule}
The decision for context $c$ at time $t$ is
\begin{equation}
T_n(c,t) \;\geq\; \tau_n(c,t) \;\;\Rightarrow\;\; \text{trust attained (grant/keep)}, \label{eq:decision}
\end{equation}
otherwise a response strategy is applied, \emph{scoped to the failing context}. For example, for a 3GPP services context, a response may be the selection of a different access stratum radio resource control encryption algorithm: a critical NF that cannot satisfy its threshold while using the strongest encryption (highest overhead) may be switched to a weaker protection algorithm as a deliberate, cost-aware trade-off, rather than being removed from the network. Responses may also adapt over time, taking feedback on which actions were appropriate in similar past events.\section{Ontology of Context-Aware Trust Management}\label{sec:ontology}

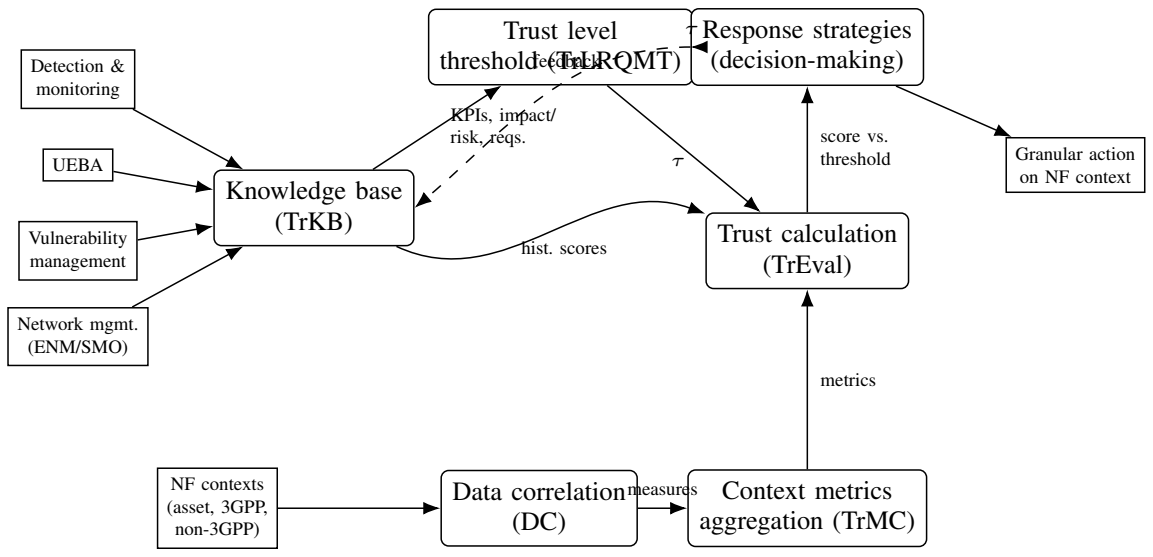
\begin{figure}[t]
\centering
\resizebox{0.85\columnwidth}{!}{%
\begin{tikzpicture}[
  comp/.style={draw, fill=white, rounded corners=2pt, align=center, font=\scriptsize, inner sep=3pt},
  ext/.style={draw, fill=white, align=center, font=\tiny, inner sep=2.5pt},
  edge/.style={-{Latex[length=1.6mm]}, font=\tiny, align=center},
  feedback/.style={-{Latex[length=1.6mm]}, font=\tiny, dashed},
  node distance=4mm and 5mm
]
\node[ext] (det) at (-3.9, 2.2) {Detection \&\\monitoring};
\node[ext] (ueba) at (-3.9, 1.4) {UEBA};
\node[ext] (vuln) at (-3.9, 0.6) {Vulnerability\\management};
\node[ext] (enm) at (-3.9, -0.2) {Network mgmt.\\(ENM/SMO)};
\node[comp] (kb) at (-1.7, 1.0) {Knowledge base\\(TrKB)};
\node[comp] (thr) at (0.6, 2.5) {Trust level\\threshold (TrLRQMT)};
\node[ext] (ctx) at (-2.6, -1.8) {NF contexts\\(asset, 3GPP,\\non-3GPP)};
\node[comp] (dc) at (0.4, -1.8) {Data correlation\\(DC)};
\node[comp] (mc) at (2.9, -1.8) {Context metrics\\aggregation (TrMC)};
\node[comp] (te) at (2.9, 0.6) {Trust calculation\\(TrEval)};
\node[comp] (rs) at (2.9, 2.5) {Response strategies\\(decision-making)};
\node[ext] (act) at (5.4, 1.4) {Granular action\\on NF context};
\draw[edge] (det) -- (kb);
\draw[edge] (ueba) -- (kb);
\draw[edge] (vuln) -- (kb);
\draw[edge] (enm) -- (kb);
\draw[edge] (kb) -- node[right, font=\tiny, align=left]{KPIs, impact/\\risk, reqs.} (thr);
\draw[edge] (thr) -- node[above, font=\tiny]{$\tau$} (rs);
\draw[edge] (thr) -- node[left, font=\tiny, pos=0.6]{$\tau$} (te);
\draw[edge] (kb) to[out=-25,in=160] node[below, font=\tiny, pos=0.55]{hist. scores} (te);
\draw[edge] (ctx) -- (dc);
\draw[edge] (dc) -- node[above, font=\tiny]{measures} (mc);
\draw[edge] (mc) -- node[right, font=\tiny]{metrics} (te);
\draw[edge] (te) -- node[right, font=\tiny, align=left]{score vs.\\threshold} (rs);
\draw[edge] (rs) -- (act);
\draw[feedback] (rs.west) to[out=180,in=40] node[above, font=\tiny, pos=0.4]{feedback} (kb.east);
\end{tikzpicture}}
\caption{Ontology of context-aware trust management as a graph. Rounded boxes are the five components; plain boxes are external and internal elements. Solid edges are data and control flows; the dashed edge is the feedback loop through which response outcomes update the knowledge base.}
\label{fig:ontology}
\end{figure}

An ontology identifies the components and concepts of a conceptualization and the key relationships and dependencies between them. Creating an ontology for context-aware TM allowed us to identify which components are required, how they interact, and which inputs each requires. Fig.~\ref{fig:ontology} presents the ontology as an explicit graph: components are nodes, and the labeled edges are the data and control flows between them. The five main components, shown as rounded nodes, are the following.

The five main components, shown as rounded nodes, are the following. \emph{Context metrics aggregation} aggregates the metric values associated with the different contexts (3GPP service, AppSS, asset) of the evaluated NF. Metric values are derived from measures produced by the data correlation component (DC), which collects data from the NFs periodically (pull) or event-triggered (push), correlates and structures it by NF identity and context, and normalizes unstructured data (\eg{} log files, library versions) into numerical values; both components can be implemented inside the NF or centrally in a security orchestrator. \emph{Trust calculation} calculates the trust score per context per NF from the aggregated metric values, following \eqref{eq:contextscore}--\eqref{eq:overallscore} or a more elaborate algorithm; it receives the current threshold from the trust level threshold component and the historical scores from the knowledge base, and produces the comparison outcome (score versus threshold). \emph{Trust level threshold}, which can also be seen as a sub-component of trust calculation, compares the calculated score against a trust level threshold and updates it according to \eqref{eq:threshold}. Its inputs include performance metrics and KPIs (MTTD, MTTR, MTTA, MTBF), impact and risk assessment (human cost, data loss, reputation damage, financial damage, system downtime, asset criticality), and historical trust scores; it may comprise one or more models, including machine-learning models, that continuously infer and recommend new thresholds per NF per context. \emph{Response strategies} acts on the comparison outcome and selects the response, applying \eqref{eq:decision}; it adapts according to historical events, taking feedback on which response actions were appropriate for a given trust score, and produces decisions actuated on the evaluated resources automatically or with human operator approval. Finally, the \emph{knowledge base} (TrKB) is a knowledge-based system component that captures the knowledge of human experts and of the network itself~\cite{kg}. It ingests and stores relevant information (\eg{} historical trust scores, security metrics and KPIs, new security and confidence-level requirements, network cost requirements, topology), models it as graphs and ontologies, and uses an inference engine with machine learning or machine reasoning models to infer relationships between data points, update trust thresholds, and predict the trust state of an NF; it can be implemented locally in an NF or centrally in a security orchestrator.

As the graph shows, the components depend on external elements (detection functions, UEBA, vulnerability management, network management), internal elements (contexts, measures, metrics), and on each other: trust calculation takes input from both context metrics aggregation and trust level threshold, which in turn takes input from the knowledge base; trust calculation influences response strategies; and response outcomes feed back into the knowledge base. The ontology is a blueprint not specific to the RAN domain: one can build on it by adding components that reflect the analyzed domain.

\section{Sequence Flow}\label{sec:flow}

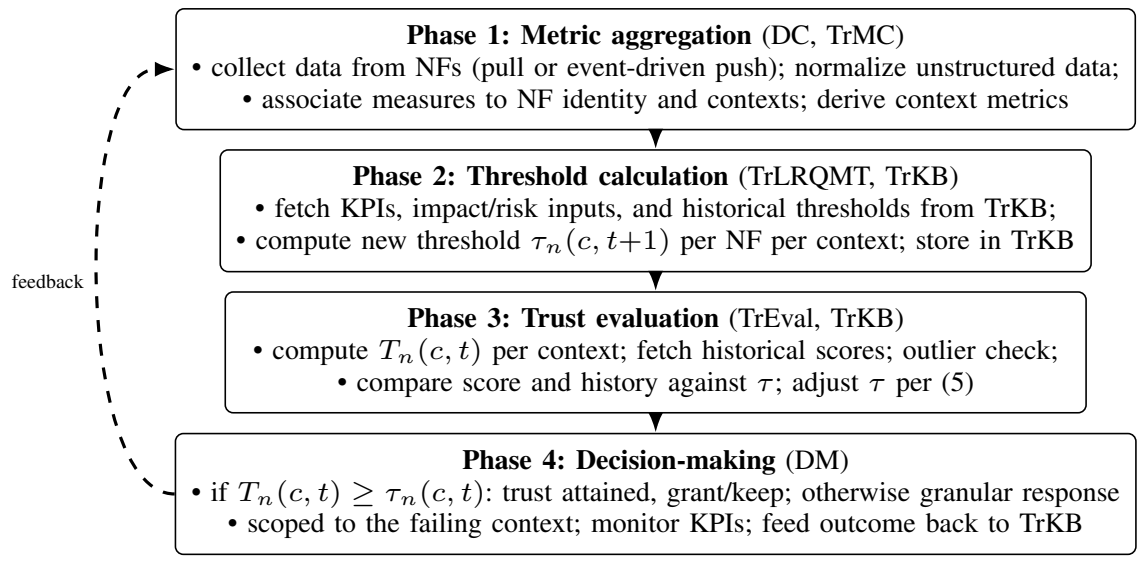
\begin{figure}[t]
\centering
\resizebox{0.85\columnwidth}{!}{%
\begin{tikzpicture}[
  phase/.style={draw, fill=white, rounded corners=2pt, align=center, font=\scriptsize, inner sep=4pt, minimum width=7.6cm},
  edge/.style={-{Latex[length=1.8mm]}, thick},
  fbedge/.style={-{Latex[length=1.8mm]}, dashed, thick}
]
\node[phase] (p1) at (0,0) {\textbf{Phase 1: Metric aggregation} (DC, TrMC)\\
\stepitem collect data from NFs (pull or event-driven push); normalize unstructured data;\\
\stepitem associate measures to NF identity and contexts; derive context metrics};
\node[phase] (p2) at (0,-1.25) {\textbf{Phase 2: Threshold calculation} (TrLRQMT, TrKB)\\
\stepitem fetch KPIs, impact/risk inputs, and historical thresholds from TrKB;\\
\stepitem compute new threshold $\tau_n(c,t{+}1)$ per NF per context; store in TrKB};
\node[phase] (p3) at (0,-2.5) {\textbf{Phase 3: Trust evaluation} (TrEval, TrKB)\\
\stepitem compute $T_n(c,t)$ per context; fetch historical scores; outlier check;\\
\stepitem compare score and history against $\tau$; adjust $\tau$ per \eqref{eq:threshold}};
\node[phase] (p4) at (0,-3.75) {\textbf{Phase 4: Decision-making} (DM)\\
\stepitem if $T_n(c,t) \geq \tau_n(c,t)$: trust attained, grant/keep; otherwise granular response\\
\stepitem scoped to the failing context; monitor KPIs; feed outcome back to TrKB};
\draw[edge] (p1) -- (p2);
\draw[edge] (p2) -- (p3);
\draw[edge] (p3) -- (p4);
\draw[fbedge] (p4.west) .. controls +(-0.9,0) and +(-0.9,0) .. node[left, font=\tiny]{feedback} (p1.west);
\end{tikzpicture}}
\caption{Sequence flow of the components of context-aware trust management, in four phases. The dashed edge is the feedback loop from decision outcomes to the next aggregation cycle.}
\label{fig:flow}
\end{figure}

Fig.~\ref{fig:flow} illustrates how the components interact to produce the output (trust scores, decisions, and actions), in four phases.

In \emph{phase 1}, data is collected, aggregated, and forwarded to the proper components. In \emph{phase 2}, the current threshold is saved in the knowledge base and a new one is calculated from the performance KPIs and impact assessments, then relayed to the trust evaluation. In \emph{phase 3}, the trust score is calculated per context and compared against the new threshold together with the historical scores; if past scores have been low on average and the current score is relatively high (or vice versa), the current score is re-evaluated, since it is likely an outlier. The comparison yields one of three outcomes: above the threshold (trust achieved), equal to it (no change), or below it (the threshold may be decreased to match the asset's actual performance). In \emph{phase 4}, the decision-making component selects the response: if the score meets the threshold, trust is attained and associated actions follow (\eg{} access granted); otherwise countermeasures are taken, \eg{} a granular action on the failing context as in Section~\ref{sec:model}. The component monitors system KPIs to evaluate whether previous responses were effective, and feeds the outcome back into the knowledge base.

\section{Deployment Options}\label{sec:deployment}

\begin{figure}[t]
\centering
\resizebox{0.85\columnwidth}{!}{%
\begin{tikzpicture}[
  mod/.style={draw, fill=white, rounded corners=1.5pt, align=center, font=\tiny, inner sep=2pt, minimum height=5mm},
  big/.style={draw, fill=white, rounded corners=2pt, inner sep=4pt},
  nf/.style={draw, fill=white, align=center, font=\tiny, inner sep=2.5pt, minimum height=5mm},
  edge/.style={-{Latex[length=1.4mm]}, font=\tiny}
]
\node[big, minimum width=7.8cm, minimum height=1.75cm] (orch) at (0,0) {};
\node[font=\scriptsize, anchor=north] at (0,0.8) {(a) Centralized: security orchestrator (e.g., ESM)};
\node[mod] (dca) at (-2.7,-0.35) {DC};
\node[mod] (tmca) at (-1.35,-0.35) {TrMC};
\node[mod] (tea) at (0,-0.35) {TrEval};
\node[mod] (dm) at (1.35,-0.35) {DM};
\node[mod] (kb) at (2.85,-0.35) {TrKB};
\node[mod] (lrq) at (2.85,-1.1) {TrLRQMT};
\draw[edge] (lrq) -- (kb);
\draw[edge] (kb) to[out=135,in=45] (tea);
\draw[edge] (dca) -- (tmca);
\draw[edge] (tmca) -- (tea);
\draw[edge] (tea) -- (dm);
\draw[edge] (dm) to[out=-60,in=120] (lrq);
\node[nf, minimum width=7.8cm] (nfs) at (0,-1.75) {RAN NFs (vDU, vCU, CU-CP, CU-UP) \quad$\bullet$\quad other domains};
\draw[edge] (nfs.north) -- (orch.south);
\node[big, minimum width=3.6cm, minimum height=1.5cm] (smo) at (-1.9,-2.9) {};
\node[font=\scriptsize, anchor=north] at (-1.9,-2.15) {(b) O-RAN: SMO};
\node[mod] (kbb) at (-2.6,-3.2) {TrKB rApp};
\node[mod] (lrqb) at (-1.2,-3.2) {TrLRQMT rApp};
\draw[edge] (lrqb) -- (kbb);
\node[big, minimum width=3.8cm, minimum height=1.5cm] (ric) at (2.0,-2.9) {};
\node[font=\scriptsize, anchor=north] at (2.0,-2.15) {(b) Near-RT RIC};
\node[mod] (dcb) at (1.1,-3.2) {DC xApp};
\node[mod] (tmcb) at (2.9,-3.2) {TrMC xApp};
\node[mod] (teb) at (1.1,-3.8) {TrEval xApp};
\node[mod] (dmb) at (2.9,-3.8) {DM xApp};
\draw[edge] (dcb) -- (tmcb);
\draw[edge] (tmcb) -- (dmb);
\draw[edge] (dmb) -- (teb);
\draw[edge] (kbb.east) to[out=0,in=180] node[above, font=\tiny]{$\tau$ stream} (teb.west);
\node[nf, minimum width=7.8cm] (nfsb) at (0,-4.5) {RAN NFs \quad$\bullet$\quad multiple rApp-suite instances per RAN network area};
\draw[edge] (nfsb.north) to[out=90,in=-90] (ric.south);
\draw[edge] (nfsb.north) to[out=90,in=-90] (smo.south);
\end{tikzpicture}}
\caption{Deployment options. (a) All components centralized in a security orchestrator collecting data from the NFs and other domains. (b) O-RAN realization: TrKB and TrLRQMT as rApps in the SMO; DC, TrMC, TrEval, and DM as xApps in the Near-RT RIC for short automation loops.}
\label{fig:deployment}
\end{figure}
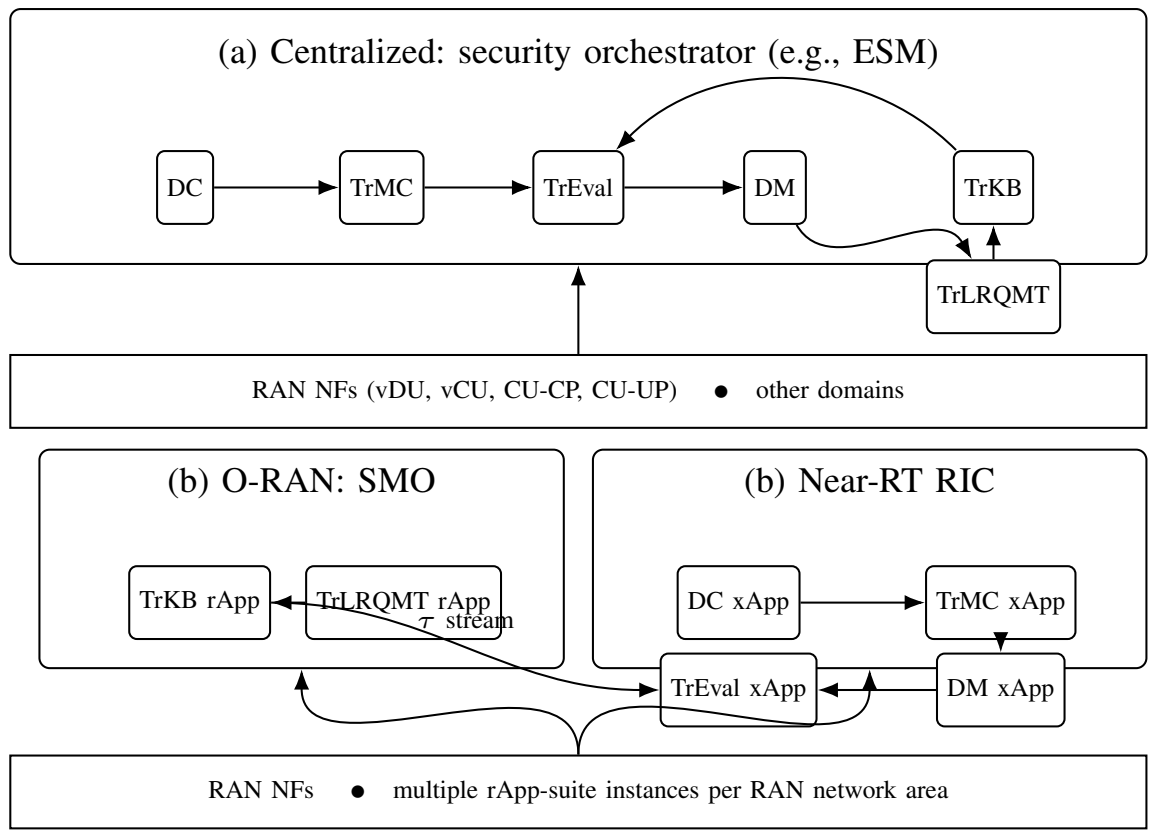

Because the framework is modular, it can be deployed in different ways depending on architectural, performance, and scalability requirements; in an autonomous network, the deployment also determines where the trust evaluations feed the closed automation loops. Fig.~\ref{fig:deployment} shows two representative options.

\subsection{Centralized and Distributed Deployments}
In the centralized option, all components are deployed in a centralized entity, \eg{} a security orchestrator such as ESM, which collects relevant data from the evaluated resources (\eg{} cloud NFs such as vCU and vDU, or O-RAN functions) and from non-security domain services, including business requirements (cost, capex, opex), topology information, and configuration information, all relevant for evaluating trust and updating the trust level requirements of each context. In the distributed option, some components are self-contained within the evaluated resource: data collection and correlation, trust metrics calculation, and even a local trust knowledge base can run locally in the NFs, while trust evaluation, trust level requirements, and decision-making remain in the centralized orchestrator with a network-wide knowledge base, improving scalability and enabling faster local decisions.

\subsection{O-RAN Realization}
In an O-RAN deployment the components can be realized as rApps or xApps. In the centralized variant, all components are rApps deployed in the service management and orchestration (SMO) framework, forming an \emph{rApp suite}: each rApp has its own distinct functionality, and used in conjunction they form the context-aware TM function. Other domains, such as security orchestration and management (SOM, \eg{} ESM or a SOC), can send additional contextual information that supplements the data collected in the knowledge base rApp. Since multiple RAN network areas (RNAs) can exist, several instances of the rApp suite can run in parallel, one per RNA. In the distributed variant, the data collection and correlation, trust metrics calculation, trust evaluation, and decision-making components run as xApps in the near-real-time RIC, closer to the NFs, enabling shorter automation loops and faster decisions with a focused view of the managed NFs; the knowledge base and trust level requirements remain rApps in the SMO, since they require a network-wide view, and the continuously updated thresholds are streamed to the trust evaluation xApp, whose decisions feed the closed automation loops of the autonomous network.\section{Discussion}\label{sec:discussion}

\subsection{Relation to Zero Trust and Autonomous Networks}
The framework operationalizes the continuous monitoring and evaluation tenets of the NIST zero-trust architecture~\cite{nist800207} for telecom networks. It also addresses a need specific to autonomous networks: the higher the autonomy level, the more decisions are taken by closed automation loops instead of human operators, and the more those decisions depend on a reliable, quantified measure of how trustworthy each function, service, and data source is~\cite{tmforum,etsian}. Context-aware TM provides exactly this measure, per context and over time, and its responses give the closed loops a safe, granular way to act on degraded trust instead of escalating every case to a human. Standardization work is ongoing, mainly in 3GPP SA3, to map the tenets of zero trust to the 5G core network~\cite{ts33501}; since trust and trustworthiness are core tenets of zero trust, there may be an opportunity to standardize \emph{what} to measure in order to calculate trust, for 5G and the AI-native networks of 6G.

\subsection{Complementarity and Generality}
Context-aware TM does not replace any existing security control. It is a process that enables an administrator, orchestrator, or automation function to take informed, risk-based decisions, and it therefore complements existing controls by requiring their input: detection functions, monitoring agents, UEBA, and vulnerability management all enrich the context and improve evaluation accuracy. Although the terminology refers to the RAN domain and O-RAN deployments, the framework applies to other RAN deployments (\eg{} gNB, eNB), to other domains (core network, transport, management), and to different mobile generations, from 4G and 5G to 6G; the context definitions of Section~\ref{sec:contexts} can be extended where the target environment requires it.

\subsection{Limitations and Future Work}
The trust calculation and response strategies presented here are deliberately simple instantiations of the framework; learning the weights $\alpha_{c,k}$ and $w_c$, designing the response policy (\eg{} with reinforcement learning in the decision-making component), and defining the measurement methods for the metrics of Table~\ref{tab:metrics} are separate studies; the ``measurable security'' work item of this research line addresses the metrics in depth. A quantitative evaluation on a live or emulated RAN testbed, including attack scenarios such as signaling storms against a DU, is a natural next step, as is studying how trust scores should gate the autonomy levels of closed automation loops.
\section{Conclusion}\label{sec:conclusion}

Context provides a granular view of the system and enables prioritized, context-appropriate decisions at a given point in time, which is especially important in a complex, dynamic, and distributed environment like the RAN, and which becomes a precondition for autonomy: a network can only automate what it can trust. This paper defined context-aware trust management for autonomous networks: two context categories (asset and services), candidate security metrics and measures, an ontology in the form of an explicit graph over five modular components, a simple trust evaluation and threshold-adaptation model, the end-to-end sequence flow, and three deployment options. Because trust evolves over time and varies across contexts, not all trust scores should be treated equally; machine reasoning in the form of knowledge-base agents allows the system to decide, at a given point in time, whether a request should be trusted. We believe this blueprint gives researchers and system architects a clear starting point for context-aware trust management in telecom networks, from 5G toward 6G.

\balance


\begin{thebibliography}{14}

\bibitem{nistglossary}
NIST, ``Trust,'' in \emph{NIST Glossary}, Computer Security Resource Center, csrc.nist.gov.

\bibitem{nistglossaryctx}
NIST, ``System context,'' in \emph{NIST Glossary}, Computer Security Resource Center, csrc.nist.gov.

\bibitem{nist800207}
S.~Rose, O.~Borchert, S.~Mitchell, and S.~Connelly, \emph{Zero Trust Architecture}, NIST Special Publication 800-207, 2020.

\bibitem{dey}
A.~K. Dey and G.~D. Abowd, ``Towards a better understanding of context and context-awareness,'' in \emph{Proc. CHI 2000 Workshop on Context-Awareness}, The Hague, Netherlands, 2000.

\bibitem{iotstudy}
``A comprehensive study on the trust management techniques in the Internet of Things,'' \emph{IEEE Xplore}, Doc. 8789637, Aug. 2019.

\bibitem{cloudsurvey}
M.~B. Monir \emph{et al.}, ``Trust management in cloud computing: A survey,'' in \emph{Proc. IEEE 7th Int. Conf. Intelligent Computing and Information Systems (ICICIS)}, Cairo, Egypt, 2015.

\bibitem{bonatti}
P.~Bonatti, D.~Olmedilla, C.~Duma, and N.~Shahmehri, ``An integration of reputation-based and policy-based trust management,'' 2005.

\bibitem{meloudi}
Y.~Meloudi \emph{et al.}, ``Trust management framework for containerized workloads: Applications to 5G networks,'' in \emph{Trust, Privacy and Security in Digital Business (TrustBus)}, Lecture Notes in Computer Science, vol. 14184, Springer, 2023, pp. 287--302.

\bibitem{kg}
A.~Hogan \emph{et al.}, ``Knowledge graphs,'' \emph{ACM Computing Surveys}, vol.~54, no.~4, pp. 1--37, 2021.

\bibitem{ts33501}
3GPP, \emph{TS 33.501: Security architecture and procedures for 5G System}, Rel.~17, 2023.

\bibitem{tmforum}
TM Forum, ``Autonomous networks: Empowering digital transformation,'' Whitepaper, 2019.

\bibitem{etsian}
ETSI, ``Autonomous networks, supporting tomorrow's ICT business,'' ETSI White Paper No.~40, 2019.

\bibitem{pat1}
``Apparatus, method and computer program product for trust management,'' U.S. Patent 11~012~313~B1, 2021.

\bibitem{pat2}
``Method and apparatus for remote trust management for machine to machine communications,'' U.S. Patent App. 2013/0097317~A1, 2013.

\end{thebibliography}
\end{document}